\documentclass[12pt]{article}

\usepackage{amsmath,amsfonts}
\usepackage{amssymb}
\usepackage{epsfig,psfrag,cite}
\usepackage[usenames,dvips]{color}
\usepackage{comment}
\usepackage{psfrag}
\usepackage{fullpage}

\makeatletter
\@addtoreset{equation}{section}
\makeatother

\newcommand{\be}{\begin{equation}}
\newcommand{\ee}{\end{equation}}
\newcommand{\bea}{\begin{eqnarray}}
\newcommand{\eea}{\end{eqnarray}}
\newcommand{\Tr}{\operatorname{Tr}}

\newcommand{\diag}{\operatorname{diag}}

\begin{document}

\thispagestyle{empty}

\begin{center}
\hfill UAB-FT-699

\begin{center}

\vspace{.5cm}

{\LARGE\bf  The Least Supersymmetric Standard Model}

\end{center}

\vspace{1.cm}

{\bf Antonio Delgado$^{\,a}$ 
and Mariano Quir\'os$^{\,a,b}$}\\

\vspace{1.cm}
${}^a\!\!$ {\em {Department of Physics, University of Notre Dame\\Notre Dame, IN 46556, USA}}

\vspace{.1cm}

${}^b\!\!$ {\em {Instituci\'o Catalana de Recerca i Estudis  
Avan\c{c}ats (ICREA) and\\ Institut de F\'isica d'Altes Energies, Universitat Aut{\`o}noma de Barcelona\\
08193 Bellaterra, Barcelona, Spain}}

\end{center}

\vspace{0.8cm}

\centerline{\bf Abstract}
\vspace{2 mm}
\begin{quote}\small
We propose the minimal (Least) version of the Supersymmetric Standard Model which can solve the hierarchy problem in the same way as the so-called Minimal Supersymmetric Standard Model (MSSM) and presents solutions to some of its problems. Supersymmetry is broken in a secluded sector and mediated to the observable sector by messengers of a gauge group $G$ under which the first two generations transform. The group $G$ spontaneously breaks (almost) supersymmetrically at a scale at most a few orders of magnitude below the scale of gauge messengers $M_*\sim 10^{15}$ GeV. By gauge mediation sfermions of the first two generations acquire supersymmetry breaking masses $\widehat m \sim 10$ TeV. Supersymmetry breaking is also mediated by gravity which generates masses for all sfermions, Higgsinos and gauginos at the TeV scale and can provide appropriate values to the  $\mu$ and $B_\mu$ parameters by $D$-term effective operators. If gravity mediation is Minimal Flavor Violating there is no supersymmetric flavor problem. In the presence of $R$-parity dark matter can be the lightest neutralino, as in the MSSM, and the LHC model phenomenology is characterized by the fact that only third generation squarks and sleptons are present.
\end{quote}

\vfill

 \newpage

\textit{1. Introduction.} 
Supersymmetry, as a perturbative theory valid up to very high scales, remains the most appealing solution to the Standard Model (SM) hierarchy problem. In particular its $R$-parity conserving minimal supersymmetric extension (MSSM) with all supersymmetric partners at the TeV scale provides gauge coupling perturbative unification at scales $M_G\sim 2\times 10^{16}$ GeV, high enough not to produce visible proton decay processes although far enough from the Planck scale $M_P\sim 2.4\times 10^{18}$ GeV for quantum gravitational effects to be negligible, and candidates for Dark Matter in the gaugino-Higgsino sector. In view of present searches and bounds at the LHC~\cite{Atlas,CMS} it is interesting to reappraise the minimal set of parameters in the MSSM at the TeV scale which are consistent with theoretical and phenomenological constraints.

The first consideration is fine-tuning. In fact although a stop squark sector not heavier than the TeV scale appears to be consistent in the MSSM with present bounds on the Higgs mass~\cite{Carena:1995wu} (which admittedly generates a little hierarchy problem) this is not so for the first and second generation squarks which are much more weakly coupled to the Higgs. While, in view of present bounds on squark masses from LHC, we are not trying to solve here the MSSM little hierarchy problem one can easily increase the mass of the first and second generation by one order of magnitude without worsening the fine-tuning as we will see next, which will completely change the supersymmetric phenomenology at the LHC.

The second consideration is the supersymmetric flavor problem which imposes very strong constraints on the masses of first and second generation squarks unless there is some alignment or mass degeneracy mechanism. Of course this problem depends on the mechanism by which supersymmetry breaking is communicated from the hidden to the observable sector. In particular the supersymmetric flavor problem can be solved if supersymmetry breaking is transmitted by gauge interactions, the so-called gauge mediated supersymmetry breaking  (GMSB) mechanism~\cite{gmsb}. However this mechanism fails to solve in a simple way the so-called $\mu/B_\mu$ problem, which is at the basis of electroweak symmetry breaking, and does not generate any $A$-term which requires heavier stops and makes more acute the little hierarchy problem. Moreover supersymmetry breaking implies a non-zero gravitino mass and gravitational interactions will transmit supersymmetry breaking to the observable sector, the so-called gravity mediation supersymmetry breaking (GrMSB)~\cite{nilles}, while if the gravitino mass is at the electroweak scale they can provide a natural solution to the $\mu/B_\mu$ problem by generating non renormalizable operators involving the Higgs sector~\cite{giudicemasiero}. However from the point of view of the effective theory there is no reason why gravitational interactions should be flavor blind and in principle they could create a severe supersymmetric flavor problem if GrMSB is not small enough with respect to GMSB. In the MSSM, where all GMSB masses are at the TeV scale, solving the supersymmetric flavor problem would require a small enough gravitino mass which would jeopardize electroweak symmetry breaking. 

In this paper we propose a simple model where the above problems are easily solved. In particular supersymmetry breaking is communicated to the first and second generation sfermions by gauge interactions (at some scale higher than the TeV but still consistent with the fine-tuning considerations) while gravity communicates supersymmetry breaking to all generations as well as to the Higgs and gauge sector at the TeV scale.  This model will share and improve the goodness of GMSB because the first and second generation are degenerate \text{and} heavy which will be of great help in suppressing FCNC operators. It will also share the goodness of GrMSB models by which the $\mu/B_\mu$ problem is solved by the Giudice-Masiero mechanism and the gravitino is heavy enough such that the lightest supersymmetric particle (LSP) can be the lightest neutralino which constitutes the most familiar component of Dark Matter. An important ingredient of the model is that it is minimal so it does not require any extra stuff or gauge symmetry at the TeV scale. In fact at the TeV scale it is a theory which contains a supersymmetric gauge, Higgs and third generation sector plus a non-supersymmetric sector with the first and second generation of fermions. In this sense it is at low energy the least supersymmetric Standard Model (LSSM). Notice that because two complete scalar generations are heavier than the third one the MSSM unification properties are kept in the LSSM with the same unification scale $M_G$ and just a tiny modification of the MSSM unification coupling $\alpha_G$.

Let us notice that a similar kind of construction, dubbed \textit{effective} supersymmetry~\cite{Cohen:1996vb}, was already proposed in the past. Although both constructions share many common features there are some essential differences. In particular the models of Ref.~\cite{Cohen:1996vb} make use of a strongly interacting gauge group to which the first and second generation couple more strongly than the third one and contain a scale of compositeness which, if below $M_G$, could be in trouble with unification. These problems will be absent in our constructions. Also more phenomenological analyses of models with \textit{non-standard} supersymmetric spectra have been recently studied~\cite{Barbieri:2010pd}, as well as in connection with the flavor problem~\cite{Barbieri:1995uv}.

\textit{2. The model.} Now we will present a particularly simple model where the first and second generation of sfermions  get a degenerate mass. We will assume as in GMSB a secluded sector where supersymmetry is broken at some scale $M_*$ by a chiral superfield $X$ via some mechanism which we will leave unspecified since it is not relevant for the discussion
\be
X=M_*+\theta^2 F
\ee
and where as usual we assume $\sqrt{F}\ll M_*$. This breaking will be communicated to the first and second generation sfermions by messengers of the gauge interactions of some group $G$ under which sfermions transform and which commute with the SM gauge interactions, which guarantees that the SM Higgs and gauge sectors do not acquire any mass. 

More precisely if we consider the three SM generations $\psi_i=(Q,L,U^c,D^c,E^c)_i$ [$i=(a,3),\,  a=(1,2)$] then $\psi_a$ should transform under $G$ while $\psi_3$ is a singlet. One possibility is that $G=SU(2)$ and $\psi_a$ transforms as a doublet~\footnote{The $SU(2)$ group can be part of a flavor symmetry introduced to explain the structure of quark masses and CKM matrix elements as in Refs.~\cite{Pomarol:1995xc,barbieridvali}.} or more simply $G=U(1)\subset SU(2)$ as for instance the $U(1)$ generated by the $T_3=\diag(1,-1)$ generator of $SU(2)$. In fact for simplicity we will consider the latter possibility with a $U(1)$ gauge group with gauge coupling $\widehat g$ and hypercharge $\widehat Y$ assignment
\be
\widehat Y_{\psi_{1,2}}=\pm 1,\quad \widehat Y_{\psi_{3}}=0
\ee
The $U(1)$ gauge theory is by construction anomaly free because the anomalies cancel between the first and the second generation. We will also introduce messengers $\Phi_{1,2}$ with hypercharges $\widehat Y_{\Phi_{1,2}}=\pm 1$ coupled to the superfield $X$ by the superpotential term
\be
W=\Phi_2 X \Phi_1
\ee

Gauge interactions mediated by the $U(1)$ gauge bosons and corresponding gauginos $(\widehat A_\mu,\widehat \lambda)$ will transmit supersymmetry breaking to the sfermions of the first and second generations and give them a common mass $m_{\widetilde Q_{1,2}}=m_{\widetilde U^c_{1,2}}=m_{\widetilde D_{1,2}^c}=m_{\widetilde L_{1,2}}=m_{\widetilde E^c_{1,2}}=\widehat m$ with~\cite{gmsb}
\be
\widehat m^2=2\frac{\widehat \alpha^2(M_*)}{16\pi^2}\,\frac{F^2}{M_*^2}
\label{mhat}
\ee
as well as a similar Majorana mass to the gaugino $\widehat\lambda$: $M_{\widehat\lambda}\simeq \widehat m$. While we will postpone a more precise constraint on $\widehat m$ we just point out that, as stated above, we will require that $\widehat m\gg 1$ TeV. 

Also notice that the $U(1)$ gauge symmetry should be spontaneously broken at some scale $v$ below $M_*$ when some (SM singlet) Higgs fields $\varphi_{1,2}$ with hypercharges $\widehat Y_{\varphi_{1,2}}=\pm 1$ acquire vacuum expectation values (VEV) along the direction $\langle \varphi_1\rangle=\langle\varphi_2\rangle=v$  not to create a $D$-term breaking mass for first and second generation sfermions. In fact the $U(1)$ gauge symmetry does forbid some Yukawa couplings which should be generated after spontaneous symmetry breaking by non-renormalizable superpotential operators as~\cite{barbieridvali}
\be
\frac{1}{M^2_*}\left(  y_{11}\varphi_2^2\, \psi_1 H\psi^c_1+y_{22}\varphi_1^2\,  \psi_2 H\psi^c_2\right)+
\frac{1}{M_*}\left(  y_{13}\varphi_2\, \psi_1 H\psi^c_3+y_{23}\varphi_1\, \psi_2 H\psi^c_3\right)
\ee
where $H$ stands for either $H_2$ or $H_1$ depending on the particular SM structure of the coupling. In particular these operators can be generated by integration of massive vector like scalar fields with a renormalizable superpotential as in Ref.~\cite{barbieridvali}. Although the precise value of $v/M_*$ will depend on the particular theory describing the flavor in the quark sector one can generically deduce that $v$ should be at most a few orders of magnitude below $M_*$.
In fact let us notice that although the $U(1)$ symmetry should not be identified with a flavor symmetry it can be embedded into it and should not forbid some Yukawa couplings as e.g.~$Y_{23}^{U,D}$. In particular the most stringent condition comes from the hierarchical structure of the fermion mass matrix~\cite{Hall:1993ni} in the up sector which yields
%
$Y^U_{23}\equiv v\,y^U_{23} /M_*\simeq \sqrt{m_cm_t/v_U^2 }\simeq 10^{-1}$
%
[where $v_U$ ($v_D$) stands for the VEV of $H_2$ ($H_1$)] which, assuming that the couplings $y_{ij}^{U,D}$ stay in perturbative values, puts the lower bound $v/M_*\gtrsim 10^{-2}$. On the other hand in the leptonic sector the right handed neutrino supermultiplets $N^c_i$ are $U(1)$ singlets and any structure for the Majorana mass matrix determined by the corresponding flavor symmetry will be allowed by the $U(1)$ gauge symmetry. 

A simple mechanism to spontaneously break the $U(1)$ symmetry is by the superpotential
\be
W=\lambda S(\varphi_1\varphi_2-v^2)
\ee
where a $U(1)$ singlet field $S$ has been introduced. Even if the fields $\varphi_{1,2}$ acquire by gauge mediation a supersymmetry breaking mass $\widehat m$ as in Eq.~(\ref{mhat}) since the scale of $U(1)$ breaking is $v\gg \widehat m$ we will  safely neglect for the moment the latter and consider the supersymmetric breaking of $U(1)$. The supersymmetric potential is then
\be
V_{SUSY}=\lambda^2\left| \varphi_1\varphi_2-v^2\right|^2+\frac{\widehat g^2}{2}(|\varphi_1|^2-|\varphi_2|)^2+\lambda^2|S|^2(|\varphi_1|^2+|\varphi_2|^2)
\label{susypot}
\ee
whose minimization yields $\langle S\rangle=0$, $\langle\varphi_1\rangle=\langle\varphi_2\rangle=v$ and the spectrum consists in a massive gauge vector multiplet $(\widehat A_\mu, Re(\varphi_1-\varphi_2),\widehat\lambda, \tilde\varphi_1-\tilde\varphi_2)$ with a mass $\sqrt{2}\widehat g v$ and a massive chiral multiplet $(S,\varphi_1+\varphi_2,\tilde S,\tilde\varphi_1+\tilde\varphi_2)$ with a mass $\sqrt{2}\lambda v$. 

Of course the gauge mediation mechanism gives a common supersymmetry breaking soft square mass $\widehat m^2$ to $\varphi_{1,2}$ which translates into a tiny modification in the previously obtained supersymmetric potential (\ref{susypot}) as 
\be
V_{SOFT}=\widehat m^2(|\varphi_1|^2+|\varphi_2|^2).
\ee
Its minimization translates in particular into the shift $\langle\varphi_1^2\rangle=\langle\varphi_2^2\rangle\equiv \widehat v^{\,2}=v^2-\widehat m^2/\lambda^2$ while the supersymmetric spectrum is spoiled by $\mathcal O(\widehat m^2/v^2)$. In particular there are two scalars, $[Re\, S,Re\,(\varphi_1+ \varphi_2)]$,  with degenerate masses, $\sqrt{2}\lambda \widehat v$ and one scalar, $Re(\varphi_1- \varphi_2)$, with square mass $2\widehat g^2\widehat v^2+\widehat m^2$. There are also two pseudoscalars $[Im\, S,Im\,(\varphi_1+\varphi_2)]$ with degenerate masses $\sqrt{2}\lambda v$. In the fermionic sector there are two degenerate Weyl spinors $(\widetilde S,\widetilde\varphi_1+\widetilde\varphi_2)$ with masses $\sqrt{2}\lambda \widehat v$, while  the Weyl fermion $\widetilde\varphi_1-\widetilde\varphi_2$ and the gaugino $\widehat \lambda$ get mixed with mass eigenvalues 
%
$
M_{\pm}=
\sqrt{2}\widehat g v\pm \frac{1}{2}M_{\widehat\lambda}+\mathcal O(M_{\widehat\lambda}^2/\widehat g^2 v^2)
$.

For the moment we have not broken supersymmetry neither in the SM gauge and Higgs sectors nor in the third generation of quarks and leptons sector. However gravity is a universal messenger of supersymmetry breaking and in general it cannot be neglected neither in the sectors where supersymmetry is unbroken nor in the sector where supersymmetry was already broken by gauge interactions since it can create flavor problems. In fact in any supergravity theory supersymmetry breaking appears with a non-vanishing gravitino mass which, from general arguments based on the cancellation of the cosmological constant, is given by
\be
m_{3/2}\simeq\frac{F}{\sqrt{3}M_P}
\label{gravitinomass}
\ee
where the numerical prefactor  is theory dependent and we will consider generically to be $\mathcal O(1)$. The main drawback of GrMSB as the only source for communication of supersymmetry breaking is precisely that there is no generic reason why it should be flavor blind, unlike the GMSB mechanism. In principle it will provide supersymmetry breaking masses $m_{ij}^2$ and trilinear couplings $A_{ij}^{U,D}$, which are not necessarily flavor diagonal~\footnote{Unless there is some flavor symmetry in the underlying supergravity or string theory.}, on top of the gaugino masses $M_A$. In the absence of a particular fundamental underlying theory one can assume that those masses are generated from effective operators as in 
\be
\frac{1}{M_P^2}\int d^4\theta XX^\dagger Q_i^\dagger Q_j,\quad \frac{1}{M_P}\int d^2\theta XQ_iH_2 U^c_j,\quad 
\frac{1}{M_P}\int d^2\theta XW^AW^A
\ee
with $\mathcal O(1)$ coefficients, which yield that all of them are of the order $m_{3/2}$. Moreover the effective operators involving the Higgs sector
\be
\int d^4\theta X^\dagger H_1 H_2,\quad \int d^4 X^\dagger X (H_1H_2+h.c.)
\ee
provide a simple explanation~\cite{giudicemasiero} of the generation of $\mu\simeq m_{3/2}$ and $B_\mu\simeq m_{3/2}^2$ terms. Of course to get a realistic theory of electroweak symmetry breaking $m_{3/2}$ has to be at the electroweak scale.

The GrMSB mechanism generates supersymmetry breaking parameters at the scale $Q\simeq M_P$. The corresponding parameters at the electroweak scale are obtained by integrating a set of renormalization group equations. Since we are not considering a particular supergravity model and consequently we can not make detailed predictions of the low energy supersymmetric parameters, for the purpose of this letter it is enough to consider the contribution from the dominant color $SU(3)$ corrections which are given by $m^2_{\widetilde Q_3}\simeq m^2_{\widetilde U^c_3}\simeq m^2_{\widetilde D^c_3}\simeq m_{3/2}^2+\Delta m^2$ with~\cite{gmsb}
\be
\Delta m^2=\frac{2C_3}{b_3}\left(1-\frac{\alpha_3^2(m_Z)}{\alpha_3^2(M_P)}\right)M_3^2
\label{Delta}
\ee
where $C_3=4/3$ and $b_3=-3$ are respectively the quadratic Casimir of quarks and beta coefficient for $SU(3)$ and $M_3=M_3(M_P)$ is the gluino mass generated by GrMSB. From Eq.~(\ref{Delta}) and assuming $M_3\simeq m_{3/2}$ one obtains $m_{\tilde Q_3}^2\simeq 8 m_{3/2}^2$. Of course in particular supergravity models this ratio should be computed in detail and the subsequent conclusions could change a bit although we believe that our results are rather generic. For that reason from here on we will be rather qualitative and will assume that all supersymmetry breaking parameters generated at $M_P$ are $\mathcal O(m_{3/2})$.

For the moment we have different scales, in particular $M_*$, $F$, $\widehat m$ and $m_{3/2}$ which are related to each other by phenomenological arguments. In principle the gravitino mass $m_{3/2}$ is related to the Higgs parameters $\mu$ and $B_\mu$ and to the mass of third generation squarks $m_{\tilde Q_3}$ at low scales by Eq.~(\ref{Delta}). Present bounds on the Higgs mass impose typical scales $m_{\tilde Q_3}\sim A_t\sim 1$ TeV which in turn are consistent with a gravitino mass $m_{3/2}\simeq 300$ GeV. On the other hand the gluino mass at low scales is $M_3\simeq \alpha_3(m_Z)/\alpha_3(M_G)m_{3/2}\simeq 3\, m_{3/2}\simeq 1$ TeV.

\begin{figure}[htb]
\psfrag{x}[][bl]{$\log_{10} M_*/TeV$} 
\psfrag{y}[][l]{$\widehat m$ (TeV)}
\begin{center}
\vspace{.2cm}
\includegraphics[width=0.6\textwidth]{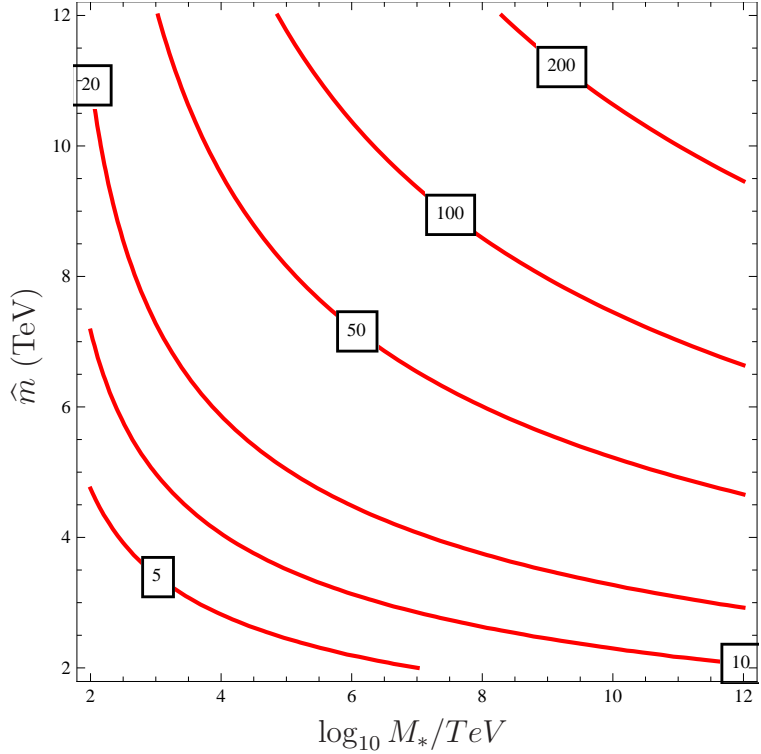}
\end{center}
\caption{\it Contour levels of the sensitivity $\Delta_{\widehat m^2}$ of $m_Z^2$ with respect to $\widehat m^2$ in the plane $(\widehat m,M_*)$. 
}
\label{sensitivity}
\end{figure}

\textit{3. The fine-tuning.} It is well known that third generation squark and gluino masses at the TeV scale generate in the MSSM a little hierarchy problem equivalent to a fine-tuning. In particular for a Higgs mass $m_H\simeq 120\, (122)$ GeV the MSSM sensitivity $\Delta$ with respect to the different parameters yields $\Delta\simeq 100\, (200)$~\cite{Chankowski:1997zh}. It is possible to alleviate (solve) this problem by enlarging the MSSM with new (singlet or triplet) states coupled to the MSSM superpotential Higgs sector~\cite{Espinosa:1998re}  or with new gauge interactions at low energy which can contribute by $F$ and/or $D$-terms to the Higgs mass~\cite{puneet}. Since we will be considering only the minimal supersymmetric extension of the Standard Model in this work, and in view of present bounds from LHC~\cite{Atlas,CMS}, it should be useless to try to improve the fine-tuning triggered by the heavy first and second generation sfermions over that which already appears due to third generation squarks and gluinos. We will then impose an upper limit on $\widehat m$ by imposing an upper bound on the sensitivity with respect to $\widehat m^2$~\cite{Barbieri:1987fn}
\be
\Delta_{\widehat m^2}=\left|\frac{\widehat m^2}{m_Z^2}\frac{\partial m_Z^2}{\partial \widehat m^2}\right|
\label{Deltasen}
\ee
as $\Delta_{\widehat m^2}\lesssim 200$. In fact as $\Tr \widehat Y m^2=0$ the leading contribution of first and second generation sfermions appears at two-loop as~\cite{Martin:1993zk}
\be
\Delta\beta^{(2)}_{m^2_{H_{1,2}}}=\frac{3}{16\pi^2}\left(
\alpha_2^2 \Tr \left[3m^2_{\widetilde Q}+m^2_{\widetilde L}\right]+
\frac{\alpha_1^2}{25}\Tr\left[m^2_{\widetilde Q}+3m^2_{\widetilde L}+8m^2_{\widetilde U^c}+2m^2_{\widetilde D^c}+6m^2_{\widetilde E^c}\right]
\right)
\label{beta2}
\ee
Using the fact that as $\widehat m^2\gg M_3^2$ the renormalization of first and second generation sfermions, Eq.~(\ref{Delta}), is tiny and can be safely neglected one can easily approximate their correction between $\widehat m$ and $M_*$ from Eq.~(\ref{beta2}) as~\footnote{Below $\widehat m$ first and second generation sfermions are decoupled and they do not contribute to the $\beta$-function in (\ref{beta2}).}
\be
\Delta m_{H_{1,2}}^2\simeq \frac{6}{\pi}\left(\Delta \alpha_2+\frac{1}{33}\Delta \alpha_1\right)\widehat m^2
\label{ecuacion}
  \ee
where $\Delta\alpha_r\equiv\alpha_r(M_*)-\alpha_r(\widehat m)$ for $r=1,2$. We are neglecting in Eq.~(\ref{ecuacion}) the small correction similar to that of Eq.~(\ref{Delta}) produced by the running of $\widehat\alpha$ between $M_*$ and $v$ which we have checked to contribute by at most a few percent to the value of sfermion masses at the low scale. From this one can easily extract the sensitivity  with respect to $\widehat m$ as it is shown in Fig.~\ref{sensitivity}.
The fine-tuning of every contour line is one part in $\Delta_{\widehat m^2}$ so we will impose the region where $\Delta_{\widehat m^2}<200$ as the fine-tuning generated by the third generations squarks and gluinos at the TeV is no better. We can see that $\widehat m\lesssim 10$ TeV for any value of $M_*\lesssim M_G$. From here on we will fix $\widehat m=10$ TeV.
 
 Once we have fixed $\widehat m\simeq 10$ TeV, by fine-tuning arguments, and $m_{3/2}$, by the phenomenological requirement that the third generation squarks have masses in the TeV region, to cope with present bounds on the Higgs mass, one can determine the scale $M_*$ where supersymmetry should be broken for the first and second generation sfermions. In fact using Eqs.~(\ref{gravitinomass}) and (\ref{Delta}) one can straightforwardly obtain
 \be
 M_*\simeq \frac{\widetilde g^{\,2}}{4\pi}\ M_G\simeq 10^{15}\,\textrm{GeV}
 \ee
 where we are assuming for the last relation that $\widehat\alpha(M_*)\simeq 1/20$.

\textit{4. FCNC.} We now summarize here the main features of the model at low scale. The first and second generation sfermions are almost degenerate with supersymmetry breaking masses  $\widehat m\simeq 10$ TeV mediated by GMSB of a $U(1)$ symmetry under which they are charged. Supersymmetry breaking masses of third generation sfermions,  gauginos and Higgsinos are generated by GrMSB with $m_{3/2}\simeq 300$ GeV which translates into third generation squark and gluino masses at low energy  $M_3\sim m_{\widetilde Q_3}\simeq 1$ TeV. We can generically assume that GrMSB generates flavor violation in LL and RR sectors of the first and second generation squarks as
\be
\Delta \widehat m^2\equiv |m_{\widetilde Q_1}^2-m_{\widetilde Q_2}^2|\simeq m_{\widetilde Q_3}^2
\ee
The strongest constraint comes from the generation of the FCNC and CP violating effective operator 
\be
\frac{z_{sd}}{\Lambda^2}\left(\bar d_L\gamma^\mu s_L\right)^2
\label{zsdoperator}
\ee
where we identify $\Lambda\simeq \widehat m$. In particular the experimental value of the operator $\epsilon_K$ implies the constraint~\cite{Bona:2007vi,Isidori:2010kg}
\be
\left|\textrm{Im}\, z_{sd}^{exp}\right|\lesssim 3.4\times 10^{-9}\left(\Lambda/TeV\right)^2\simeq  3.4\times 10^{-7}
\ee
where in the last expression we have used $\Lambda\simeq\widehat m= 10$ TeV.
The coefficient $ z_{sd}$ has been computed in Ref.~\cite{Blum:2009sk} and it is given by
 \be
 \left|\textrm{Im}\, z_{sd}\right| \simeq \frac{\alpha_3^2}{54} f(m_{\tilde g}^2/m_{\widetilde Q_3}^2)
  \left( \Delta \widehat m^2/\widehat m^2 \right)^2 \sin\alpha\sin 2\gamma
\label{zsd}
 \ee
where $\gamma$ is the $CP$-violating phase, $\alpha$ is the angle  between the first and second generations in the mixing matrix in the gluino-quark-squark coupling, which is expected to be $\alpha\simeq 2\theta_c$ in the absence of any fine tuning, and the function $f(x)$ is given in Ref.~\cite{Blum:2009sk}.
In our case $x\ll 1$ and we can use the behavior $f(0)=11/8$ to write
\be
\left|\textrm{Im}\, z_{sd}\right|\lesssim 2.5\times 10^{-8}
 \label{zsdfinal}
 \ee
where we have used the \textit{worst} case or maximal $CP$-violation: $\sin 2\gamma=1$.  Similarly flavor violating operators involving the first or second generation of quarks with the third one (e.g. observables as~$\Delta m_{B_d}$) are sufficiently suppressed by the ratio $\Delta\widehat m^2/\widehat m^2\simeq 0.01$.

If GrMSB generates flavor violation in the LR sector by soft parameters as e.g. $v_{U,D}A_{ij}^{U,D}$ then it can generate chirality non-conserving flavor violating operators. Their effect is characterized by the parameters
\be
\left(\delta^{q}_{ij}\right)_{LR}\simeq \frac{v_{q}A^q_{ij}}{\widehat m^2} ,\quad q=U,D
\ee
and the strongest experimental constraints~\cite{Isidori:2010kg} come from $\epsilon_K$ [$\left(\delta^{D}_{12}\right)_{LR}^{exp}\lesssim 2\times 10^{-4}$] and from the neutron and quark EDM [$\left(\delta^{D}_{11}\right)_{LR}^{exp}\lesssim 4.7\times 10^{-6},\ \left(\delta^{U}_{11}\right)_{LR}^{exp}\lesssim 9.3\times 10^{-6} $]. In order to satisfy these bounds it is convenient to rely on the assumption that GrMSB is minimal flavor violating (MFV). In particular it is enough to assume that the $A$-terms are proportional to the Yukawa matrix: $A_{ij}^q\propto Y^q_{ij}m_{\widetilde Q_3}$. In this case and using the hierarchical structure of the quark mass matrix one can write~\cite{Hall:1993ni}
\be
\left(\delta^{D}_{12}\right)_{LR}\simeq \sqrt{\frac{m_dm_s}{\widehat m^2}}\frac{m_{\widetilde Q_3}}{\widehat m},\quad
\left(\delta^{q}_{11}\right)_{LR}\simeq\frac{m_{q}}{\widehat m}\frac{m_{\widetilde Q_3}}{\widehat m}
\ee
and the theory is safe. In the general case the bound from $\left(\delta^{D}_{12}\right)_{LR}$ can be satisfied for large $\tan\beta$ while the bound from EDM is dominated, for large $\tan\beta$, by $\left(\delta^{U}_{11}\right)_{LR}$ which imposes a strong constraint on $A_{11}$.

\textit{5. The phenomenology.} The phenomenology of the model is dominated by the fact that, at the LHC reach, only the third generation squarks and sleptons are present. This means that any signal involving the first two families of scalars will be missing, at least to first approximation. So, for example, the usual way of using multileptons to discover the decay chain of charginos and neutralinos will be reduced since the possible decays of these particles are now:
\begin{equation}
\chi'\to \chi W/Z,\;\;\;\;\chi'\to\chi h,\;\;\;\;\chi'\to \psi \tilde{\psi}
\end{equation}
where $\psi$ is either a tau lepton, top or bottom quark and $\tilde{\psi}$ is the corresponding scalar. Therefore the  signal with multijets plus missing energy is going to be very much enhanced from the one having leptons which could only come from the leptonic decays of $W$ or $Z$.

Gluinos will have to decay into stop/top or sbottom/bottom pairs, therefore the signal will be:
\begin{equation}
pp\to \tilde{g}\tilde{g}\to qq \bar{q}\bar{q} + \chi\chi
\end{equation}
where $q$ is either a top or a bottom and $\chi$ the LSP. There could also be some $W/Z$ coming from the chain decays and several analyses of this possibility already exist in the literature~\cite{Barbieri:2010pd}. Finally the LSP in this model could either be the gravitino or the lightest neutralino from the GrMSB. Since the gravitino in this model has a mass of around a few hundreds of  GeV, in order to avoid any cosmological problems associated with such a stable gravitino we will suppose that the LSP, and therefore the dark matter candidate, is coming from the GrMSB sector and will most likely be the lightest neutralino. This also solves the issue of gravitino dark matter in models where GMSB is the only source of supersymmetry breaking. 

\textit{6. Conclusions.} To summarize we have introduced the Least Supersymmetric Standard Model (LSSM) corresponding to the most minimal  supersymmetric extension of the SM which solves the hierarchy problem but does not have dangerous contributions to flavor observables. This is accomplished by having a single source of supersymmetry breaking ($F$) and two different mechanisms to mediate supersymmetry breaking to the observable sector: gauge mediation (at a scale $M_*$) and gravity mediation (at the scale $M_P$). The first two generations of scalars receive contributions ($\widehat m$) through GMSB based on an extra $U(1)$ symmetry under which only those  generations transform. It is well known that GMSB does not generate any flavor violating terms therefore this contribution is safe from any flavor problem coming from the first two generations. The rest of the spectrum, third generation scalars, Higgsinos and gauginos, get their masses through GrMSB which in principle does not respect the global symmetries of the SM and therefore could potentially contribute to FCNC and CP-violating observables. Since GrMSB will also contribute to the soft masses for the first two generations, passing the tests of flavor constrains, imposing a correct electroweak symmetry breaking and no extra fine-tuning in the Higgs mass fixes the values of the different mass scales. In particular: $\sqrt{F}\simeq 10^{10}$ GeV and $M_{*}\simeq 10^{15}$ GeV. They translate into $\widehat m\simeq 10$ TeV, a gravitino mass of a few hundreds of GeV and third generation squark and gluino masses in the TeV range. The main phenomenological consequences are that, since selectrons and smuons are very heavy, the branching ratios into multileptons are decreased and the decays involving multijets increased. The DM candidate on the other hand can be the lightest neutralino and not the gravitino as it happens in usual GMSB scenarios. 

Finally notice that the model we have presented depends on three ingredients: \textit{i) The mechanism of supersymmetry breaking}. We did not specify it in this paper but it is clear that supersymmetry can be broken dynamically and/or by some O'Raifeartaigh mechanism by some more fundamental theory in some hidden sector providing the scales $F$ and $M_*$; \textit{ii) The GMSB mechanism}. We have assumed here a $U(1)$ subset of the $SU(2)$ under which the first two generations transform as a doublet. Here there is not much freedom. We can use the whole $SU(2)$ group in which case the results would be similar to those obtained in this paper. Otherwise we can use a different anomaly-free $U(1)$ where anomalies cancel within every generation. In particular if we do not want to introduce extra fields (apart from the right-handed neutrinos) the only choice is the hypercharge $\widehat Y_i=q_i(B-L)$ where $q_{1,2}\neq 0$ and $q_3=0$. In this case the right-handed neutrino multiplet should transform, to cancel the anomalies, and the mass spectrum for squarks and sleptons are different so that the physical results can be a bit different from those presented in this model; \textit{iii) The GrMSB mechanism}. Here we were not assuming any particular mechanism for gravity mediation. However confronting the model with experimental data should require considering one particular model of GrMSB leading to some particular spectrum of supersymmetry breaking parameters at the TeV scale.
Lastly the LHC will have the ultimate word on which one of the several realizations of supersymmetry, if any, is realized in Nature.

\bigskip
\emph{Note:} At the time of submission of this article another paper appeared in the arXiv~\cite{Raman} where similar ideas are presented. 

\section*{Acknowledgments}
AD was partly supported by the National Science Foundation under grant PHY-0905383-ARRA. MQ was supported in part by the Spanish Consolider-Ingenio 2010 Programme CPAN (CSD2007-00042) and by CICYT-FEDER-FPA2008-01430.

\end{document}